\documentclass[preprint,12pt]{elsarticle}

\usepackage{amssymb}

\usepackage{amsmath}

\usepackage{bm}% bold math
\usepackage{xr}
\usepackage{caption}
\usepackage{multirow}
\usepackage{subcaption}
\usepackage{tabularx}
\usepackage{comment}
\usepackage{color}
\newcommand{\Tref}{T_{\rm{L}}}
\newcommand{\ci}{{\bm{c}_i}}
\newcommand{\vi}{{\bm{v}_i}}
\newcommand{\x}{\bm{x}}
\newcommand{\U}{\bm{u}}
\newcommand{\dx}{\delta x}
\newcommand{\ii}{i}
\newcommand{\tvd}{\rm TVD}
\newcommand{\cfl}{{\rm CFL}}
\newcommand{\euler}{{\rm i}}
\newcommand{\eq}{{\rm eq}}
\journal{XXX}

\begin{document}

\begin{frontmatter}



\title{Exploring shock-capturing schemes for Particles on Demand simulation of compressible flows}


\author[inst1]{Ehsan Reyhanian}

\affiliation[inst1]{organization={Department of Mechanical and Process Engineering, ETH Zurich},%Department and Organization
            % addressline={Address One}, 
            % city={Zurich},
            postcode={8092}, 
            state={Zurich},
            country={Switzerland}}

\author[inst1]{Benedikt Dorschner}
\author[inst1]{Ilya Karlin}

% \affiliation[inst2]{organization={Department Two},%Department and Organization
%             addressline={Address One}, 
%             city={City Two},
%             postcode={22222}, 
%             state={State Two},
%             country={Country Two}}

\begin{abstract}
In this exploratory study, 
we apply shock-capturing schemes within the framework of the Particles on Demand kinetic model to simulate compressible flows with  mild and strong shock waves and discontinuities. The model is based on the semi-Lagrangian method where the information propagates along the characteristics while a set of shock-capturing concepts such as the total variation diminishing  and weighted essentially non-oscillatory  schemes are employed to capture the discontinuities and the shock-waves. The results show that the reconstruction schemes are able to remove the oscillations at the location of the shock waves and together with the Galilean invariance nature of the Particles on Demand model, stable simulations of mild to extreme compressible benchmarks can be carried out. Moreover, the essential numerical properties of the reconstruction schemes such as their spectral analysis and order of accuracy are discussed. 
\end{abstract}


% \begin{highlights}
% \item Various shock-capturing schemes are explored and studied to be incorporated in the Particles on Demand (PonD) kinetic model.
% \item A Total Variation Diminishig (TVD) limiter function is developed based on a third-order Bspline to exploit its good conservation properties.
% \item Beside the developed TVD approach, a WENO-type interpolation scheme is also used to treat the oscillations at the discontinuities.
% \item Different compressible benchmarks ranging from mild to extreme cases are simulated.
% \end{highlights}


\end{frontmatter}


\section{Introduction}
Simulation of compressible high-speed flows have been a long-standing topic of research
in computational fluids dynamics (CFD). Various advanced numerical schemes have been
developed to resolve small-scale features of shocked flows as well as capturing the 
discontinuities. As a classical CFD contradiction, while a sufficient amount of dissipation is required for 
capturing the discontinuities, it can negatively affect resolving small 
structures \cite{pirozzoli2011numerical}.
It is crucial for a numerical scheme to maintain high order of accuracy
in  smooth parts of the solution while being able to capture discontinuities.
To this end, different classes of numerical schemes have been developed such as 
total variation diminishing (TVD) \cite{HARTEN1983357}, essentially non-oscillatory (ENO) \cite{HARTEN1987231}, 
weighted ENO (WENO) \cite{LIU1994200} and targeted ENO (TENO) \cite{FU2016333}.\\

The lattice Boltzmann method (LBM) is a modern approach in the field
of computational physics, as a recast of fluid dynamics into the kinetic
theory of designer particles and has shown a promising performance in
various regimes of fluid dynamics ranging from micro \cite{Kunert2007,Hyvaluoma2008}, multiphase \cite{Sbragaglia2006,Biferale2012,Benzi2009,MazloomiM2015} and
compressible \cite{prasianakis2008,frapolli2015,pond,wilde2020semi} to complex flows and turbulence \cite{Atif2017,dorschner2016entropic}.
The LB equation describes the evolution of the populations $f_i(\x,t)$ discretized in the velocity space through the discrete particle velocities $\ci; i=1,...,Q$ with simple rules of streaming and relaxation toward the local equilibrium $f_i^{\eq}(\x,t)$.\\

Despite the considerable success of LBM in recent decades, there exist
inherent restrictions associated with this method. The most important
is the violation of the Galilean invariance which limits the application
of LBM to low Mach numbers or incompressible flows \cite{succi2018lattice,kruger2017lattice}. However, due to
the extensive applications of compressible flows such as flows with strong
shocks and discontinuities or compressible multiphase flows, developing a
kinetic approach to model these setups is still an open field of research in the LB community.\\
There have been various attempts to overcome the insufficiencies in LB and enable simulations up to higher Mach numbers. Among those, the recently developed
”Particles on Demand for Kinetic Theory” or the so-called ”PonD” method
is noteworthy, which removes these limitations by defining adaptive sets of
microscopic velocities, leading to a Galilean-invariant scheme.
The main idea of PonD is to sample particle’s velocities
based on the local thermodynamics and velocity of the flow, which is significantly different from the conventional Guassian-Hermit sampling at the
core of LBM. This new representation of the kinetics with particles subject
to optimal gauges or reference frames leads to error-free equilibrium.\\

While the PonD kinetic theory holds the underlying essential basis, i.e. Galilean invariance,
it still requires to be equipped with proper numerical schemes to handle
high-Mach simulations. In particular, due to the off-lattice property
of this method, using interpolation is inevitable which potentially results
in oscillatory solutions. This especially concerns setups with strong shocks
and discontinuities. These oscillations might trigger negative pressures and
temperatures which will blow up the simulations eventually. In this paper, the necessary numerical tools for stable simulations of such flows
are developed.

\section{Kinetic equations}
\label{sec:pond/kinetic-eq}
In PonD, the discrete velocities are defined as
\begin{align}
\vi = \sqrt{\theta}\ci + \U,
\label{eq:pond/discrete-vel}
\end{align}
where $\theta=T/\Tref$ for an ideal gas, $T$ is the local temperature, $\Tref$ is a constant particular to each lattice known as the lattice temperature \cite{shyam2009lattices} and $\U$ is the local flow velocity. Equation~\eqref{eq:pond/discrete-vel} describes that the
peculiar velocities $\ci$ are first scaled by some definite factor of the square root of the local
temperature and then shifted by the local velocity of the flow. While the former revokes the
restriction on the lattice temperature $\Tref$, the latter results in Galilean invariance.
The populations corresponding to the reference frame $\lambda=\{T,\U\}$ are denoted by $f_i^\lambda$.
Similar to LBM, the kinetic equations can split into two main parts; Collision with an exact equilibrium populations
\begin{align}
f_i^*(\x,t) = f_i(\x,t) + \omega(\rho W_i - f_i)_{(\x,t)},
\end{align} 
where  $f_i^*(\x,t)$ are the post-collision populations which are computed at the gauge $\lambda=\lambda(\x,t)$, $\omega$ is the relaxation parameter related to the viscosity and $W_i$ are conventional LBM lattice weights known for any
set of discrete speeds $\mathcal{C}$. 
The streaming step shall be implemented via the semi-Lagrangian method where the 
information at the monitoring point $(\x,t)$ is updated by traveling back through the 
characteristics to reach the departure point $\x_d(i) = \x - \vi\delta t$. However, due to the dependency of 
the discrete velocities \eqref{eq:pond/discrete-vel} on the local flow field, the departure point may be located
off the grid points. This is in contrast to LBM, where the lattice provides exact streaming along the links. Hence,
the information at the departure point must be interpolated through the neighboring points. Furthermore, 
in order to be consistent, the populations at the departure point must be in the same reference frame as the
monitoring point. Hence, the populations at the collocation points  used for the interpolation are
first transformed to the gauge of the monitoring point and then interpolated \cite{pond}.
Finally, the advection step is formulated as
\begin{align}
f(\x,t) = \sum_{p=0}^{N-1} \Lambda(\x_d-\x_p)\mathcal{G}_{\lambda_p}^{\lambda}f^{*\lambda_P}(\x_p,t),
\label{eq:SL-advection}
\end{align}
% \ik{arguments seem to be confused ...
% removed the subscript "i"}
where $\bm{x}_p$, $p=0,...,N-1$ denote the collocation points (grid points) around the departure point and $\Lambda$ is the interpolation kernel.
As mentioned before, the populations are transformed using the transformation Matrix $\mathcal{G}$.
In general, a~set of populations at gauge $\lambda$ can be transformed to another gauge $\lambda^\prime$ by matching {the} $Q$ linearly independent moments:
\begin{align}
    \bm{M}_{mn}^\lambda = \sum_{i=1}^Q f_i^\lambda v_{ix}^m v_{iy}^n,
\label{eq:independent-moments}
\end{align}
where $m$ and $n$ are integers. This may be written in the matrix product form as $\bm{M}^\lambda = \mathcal{M}_\lambda f^\lambda$ where $\mathcal{M}$ is the $Q\times Q$ linear map. 
Requiring that the moments must be independent from the choice of the reference frame, leads to the matching condition:
\begin{align}
    \mathcal{M}_{\lambda^\prime}f^{\lambda^\prime} = \mathcal{M}_{\lambda}f^{\lambda}, 
\end{align}
which {yields} the transformed populations:
\begin{align}
    f^{\lambda^\prime} = \mathcal{G}_\lambda^{\lambda^\prime}f^\lambda=\mathcal{M}_{\lambda^\prime}^{-1}\mathcal{M}_{\lambda}f^{\lambda}.
\end{align}
\\

Finally, the macroscopic values are evaluated by taking the pertinent moments
\begin{align}
\rho &= \sum_i f_i,\\
\rho\U &= \sum_i f_i\vi,\\
\rho u^2 + D\rho T &= \sum_i f_i v_i^2,
\end{align}
The implicitness in the above equations require a predictor-corrector step  to 
find the co-moving reference frame. Hence, the same procedure is repeated by
imposing the new evaluated velocity and temperature until the convergence is 
achieved. For more details, see \cite{pond}.\\

In this paper, the ideal-gas EoS $p=\rho e(\gamma-1)$ is adopted, where $e=e(T)$ 
is the specific internal energy and the specific-heat ratio is set to $\gamma=1.4$ unless stated otherwise.
To have an arbitrary value of $\gamma$, a second set of populations is employed
\cite{karlin2013consistent}. However, when using a standard lattice such as $D2Q9$,
they are designed to carry the total energy with
the equilibrium \cite{reyhanian2020thermokinetic}
\begin{align}
g_i^{\rm eq} = \rho W_i \left(2e - DT + v_i^2\right),
\end{align}
where $D$ is the dimension.
\begin{comment}
For the 25 discrete velocities, 
the second populations simply carry the excess internal energy. The equilibrium
for this type becomes \cite{frapolli2016entropic}
\begin{align}
g_i^{\rm eq,25} = \rho W_i \left(2e - DT\right).
\end{align}
In general, we can write the equilibrium as a linear combination of both equilibria
\begin{align}
g_i^{\rm eq} = \rho W_i \left(2e - DT + (1-\phi)v_i^2\right),
\label{eq:compressible/g-equilibria}
\end{align}
where $\phi_9=0$  and $\phi_{25}=1$. Then, the energy is evaluated as\\
\begin{align}
2\rho E = 2\rho e + \rho u^2 = \phi\sum_i f_i  v_i^2 + \sum_i g_i.
\end{align}
\end{comment}

Finally, we comment that the sign $u$ is interchangeably used in this paper as the flow velocity and also the solution function.

\section{Reconstruction step}
The reconstruction step is one of the most crucial
elements in PonD during the advection process. While the
transformation part is done merely by the moment-invariance rule,
there are various options for the interpolation process. The choice of the
reconstruction scheme will strictly affect the solution as well as numerical
properties such as conservation and oscillations. In this section, we will
explore a wide range of reconstruction schemes, from the basic interpolation methods to non-oscillatory high-resolution schemes and assess their
performance in PonD. In order to have a shock-capturing scheme, we make use of the TVD (Total Variation Diminishing) principle, as well as the WENO (Weighted Essentially non-Oscillatory) method.

\subsection{Interpolation schemes}
\begin{figure}[!t]
    \centering
    \includegraphics[clip, trim = 0 0.5cm 0 0, width=0.7\textwidth]{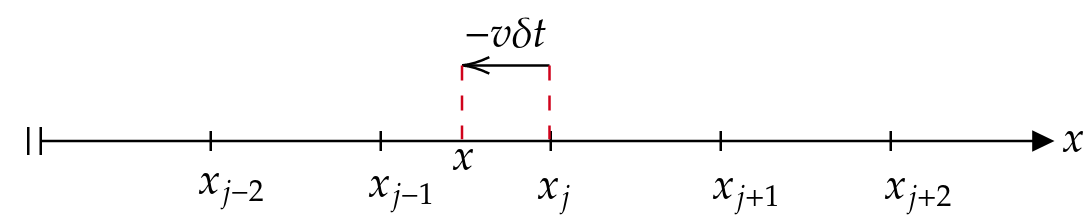}
    \caption{Schematic of the semi-Lagrangian advection in one dimension during the time step $\delta t$ along the characteristic velocity $v$.}
    \label{fig:grid}
\end{figure}

Here, we will elaborate the interpolation schemes we have used in this paper. Our experiments show that the choice of the interpolation kernel has significant effects on the accuracy and validity of the results.\\
We start by considering the one-dimensional semi-Lagrangian advection along the characteristic velocity $v$ during one time step $\delta t$. The domain is discretized into $Nx$ points $x_j;j=0,...,Nx-1$ using equally distant intervals $\dx = x_j-x_{j-1}$. Without loss of generality, we assume that $v>0$ and $x=x_j-v\delta t$ is the interpolating point (see Fig. \ref{fig:grid}). The fundamental formula for the interpolation reads \cite{Schoenberg1946}
\begin{align}
\Tilde{\phi}(x,\dx) = \sum_{j=-\infty} ^ \infty \phi(x_j)\Lambda(x-x_j,\dx),
\end{align}
where $\Lambda$ is the interpolation kernel.
\subsubsection{Lagrange polynomials}
 The basic interpolation model is the Lagrange polynomials denoted by $L_N(u)$, where $N$ is the order of interpolation and $u=|x-x_j|/\dx$. In this paper, we use the 4-point stencil centered around the interpolating point $S=\{x_{j-2},x_{j-1},x_j,x_{j+1}\}$. The resulting kernel is a $4^{\rm th}$ order accurate interpolating kernel:
 \begin{align}
 L_4(u) = \Bigg\{ \begin{tabular}{ll}
$-\frac{1}{2}(1-u^2)(u-2)$,      &  $0\leqslant u \leqslant 1$, \\
$-\frac{1}{6}(u-1)(u-2)(u-3)$,      & $1\leqslant u \leqslant 2$,\\
$0$,      & otherwise.
 \end{tabular}
 \label{eq:recons/everett}
 \end{align}
In the literature, this is also known as the Everett's formula. \cite{monaghan1985extrapolating}
\subsubsection{Moment conserving schemes}
We assume that the quantity $q$ at a set of points $x_p$ is interpolated 
through the mesh points $x_\ii$. The interpolated value of the quantity $q$ becomes \cite{rees20143d,koumoutsakos1997inviscid}
\begin{align}
q_p = \sum_\ii q_\ii \Lambda\left(\frac{x_\ii-x_p}{\dx}\right).
\label{eq:recons/M2P}
\end{align}
Similarly, one can revert the same procedure to get the values of the field $q$
 at mesh points
\begin{align}
q_\ii = \sum_p q_p \Lambda\left(\frac{x_p-x_\ii}{\dx}\right).
\label{eq:recons/P2M}
\end{align}
To conserve the first $r$ moments, the interpolation kernel must satsify the following condition
\begin{align}
\sum_\ii q_\ii (x_\ii - x)^\alpha = \sum_p q_p (x_p - x)^\alpha,
\label{eq:recons/condition}
\end{align}
where $0\leqslant \alpha < r$. It is clear that $r=0$ implies the conservation of the field $q$. 
The first $r$ moments of the field $q_\ii$ can be obtained using Eq.~\eqref{eq:recons/M2P},
\begin{align}
\sum_\ii q_\ii (x_\ii - x)^\alpha = \sum_p q_p \sum_\ii  \Lambda\left(\frac{x_p-x_\ii}{\dx}\right) (x_\ii - x)^\alpha.
\end{align}
Using the Newton formula, one can write the latter as
\begin{align}
&\sum_\ii q_\ii (x_\ii - x)^\alpha =\nonumber\\
&\sum_p q_p \sum_{k=0}^\alpha \sum_\ii 
\Lambda\left(\frac{x_p-x_\ii}{\dx}\right) x_\ii^k (-x)^{\alpha-k}
\left(\begin{tabular}{cc}
$\alpha$\\
$k$
\end{tabular}
\right).
\end{align}
Finally, we note that the latter formula can be reduced to 
\begin{align}
\sum_\ii q_\ii (x_\ii - x)^\alpha &= \sum_p q_p \sum_{k=0}^\alpha 
 x_p^k (-x)^{\alpha-k}
\left(\begin{tabular}{cc}
$\alpha$\\
$k$
\end{tabular}
\right)
\\ \nonumber
&= \sum_p q_p (x_p-x)^\alpha,
\end{align}
if and only if
\begin{align}
\sum_\ii \Lambda\left(\frac{x_p-x_\ii}{\dx}\right) x_\ii^k = x_p ^ k.
\label{eq:recons/kernel-property}
\end{align}
In other words, the property \eqref{eq:recons/kernel-property} is the necessary condition 
for the interpolation kernel $\Lambda$  to conserve the first $r$ moments, i.e. to satisfy Eq.~\eqref{eq:recons/condition}.\\
However, it can be shown that applying condition \eqref{eq:recons/kernel-property} on a 4-point stencil leads to the 
Everett's formula \eqref{eq:recons/everett}.
\subsubsection{B-Splines}
\label{sec:recons/bspline}
\begin{figure}[!t]
    \centering
    \includegraphics[clip, trim = 0 0.5cm 0 2cm, width=0.7\textwidth]{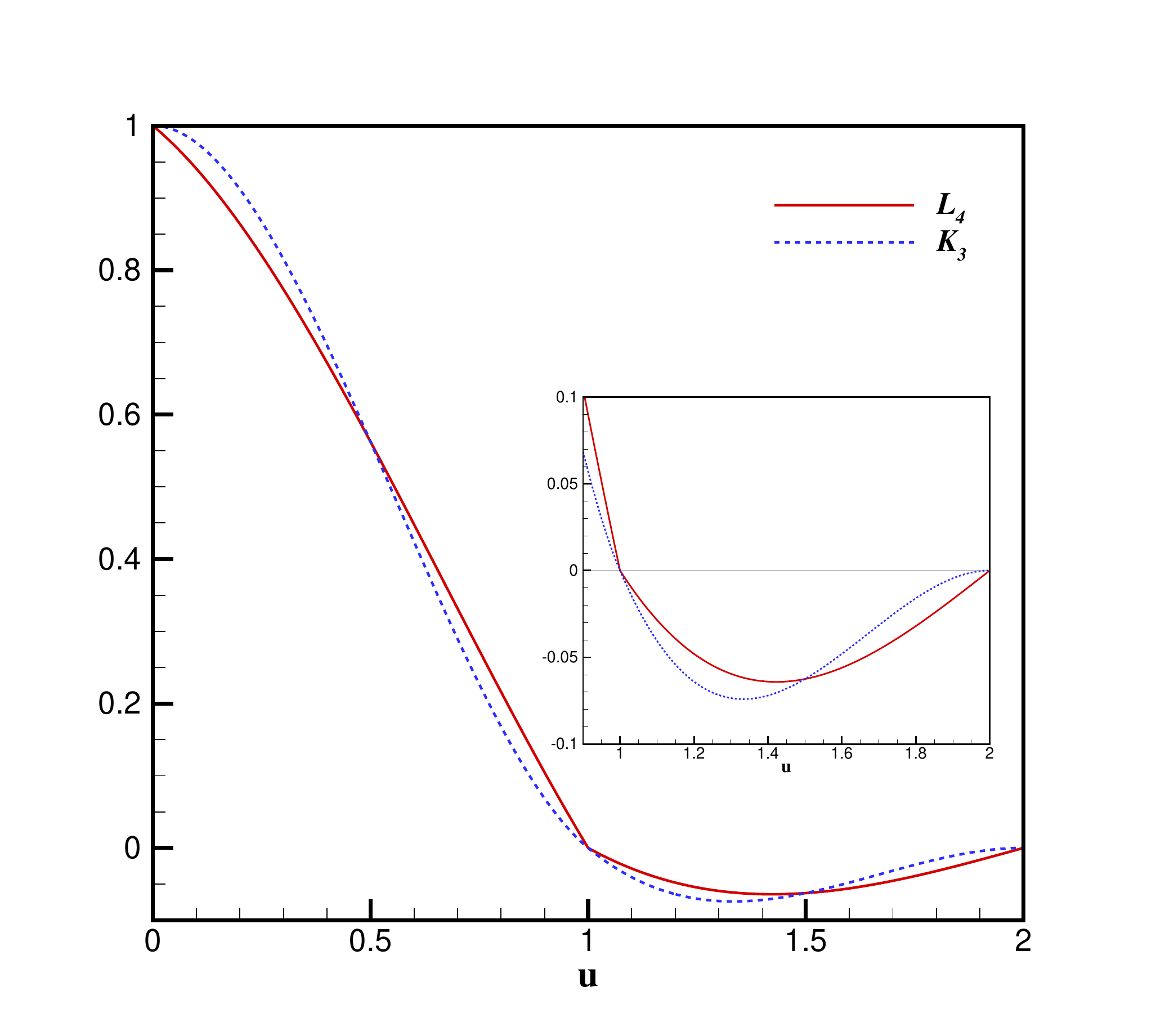}
    \caption{Comparison between $L_4$ and $K_3$ kernels. Contrary to $L_4$, $K_3$ is smooth and its derivative is defined everywhere (See inset).}
    \label{fig:kernels}
\end{figure}

It is well-known that interpolation schemes may introduce large errors when large fluctuations are present \cite{koumoutsakos1997inviscid}.
For this purpose, B-Splines are designed such that these effects are minimized \cite{Schoenberg1946}. B-Splines are non-negative functions that are generated recursively by
\begin{align}
    B^{n+1}(x)=B^n(x)*B^0(x),
\end{align}
where $*$ is the convolution operator and
\begin{align}
    B^0(x)=\Bigg\{ \begin{tabular}{ll}
    1,     &  $|x|\leqslant 1/2$,\\
    0,     & otherwise,
    \end{tabular}
\end{align}
is the nearest grid point (NGP) interpolation. The first two members of the B-Spline family fall into the category of ordinary interpolation functions, meaning $B^n(0)=1; n<2$, while the rest are
smoothing functions since $B^n(0)\neq 1; n\geqslant 2$.\\
B-Spline kernels are smooth functions and they have an accuracy of $\mathcal{O}(\dx^2)$. In ref. \cite{monaghan1985extrapolating}, it was shown that their order of accuracy can be improved for $n\geqslant2$ using Richardson extrapolation. The resulting function is,
\begin{align}
    K_n = \frac{1}{2}\left[3B_n-h\frac{\partial B_n}{\partial h}\right],
\end{align}
where $h=\dx$ is the grid spacing and the derivative of $B$ is not defined everywhere for $n<2$. Finally, reminding the notation of $u$, the third-order accurate improved B-Spline is derived as,
 \begin{align}
 K_3(u) = \Bigg\{ \begin{tabular}{ll}
$1-\frac{5}{2}u^2-\frac{3}{2}u^3$,      &  $0\leqslant u \leqslant 1$, \\
$\frac{1}{2}(2-u)^2 (1-u)$,      & $1\leqslant u \leqslant 2$,\\
$0$,      & otherwise.
 \end{tabular}
 \label{eq:recons/k3}
 \end{align}
 Both $L_4$ and $K_3$ kernels are depicted in Fig. \ref{fig:kernels}. One could observe that the $K_3$ kernel is much smoother than its counterpart and has a continuous derivative through its defined range. Also, it is visible that both kernels possess negative values which is known to cause oscillatory solutions 
in sharp contacts. We will address this issue in the followings.
% Figure \eqref{fig:recons/ovs} shows the results of the order verification study (OVS) of both kernels. The results are the outcome of the semi-Lagrangian solution of the advection equation initialized with a Guassian profile. The CFL number is fixed at $0.1$. As expected, the underlying order of accuracy of both schemes is recovered.
% \begin{figure}[!t]
%     \centering
% \begin{subfigure}{\linewidth}
% \centering
%     \includegraphics[clip, trim= 0 0 0 2cm, width=\textwidth]{OVS-L4}
%     \caption{$L_4$ kernel}
%     \label{ovs-L4}
% \end{subfigure}
% \vfill
% \begin{subfigure}{\linewidth}
% \centering
%     \includegraphics[width=\textwidth]{OVS-K3}
%     \caption{$K_3$ kernel}
%     \label{ovs-k3}
% \end{subfigure}
% \caption{OVS of $L_4$ and $K_3$ interpolation schemes. The results are obtained by solving the advection equation at $CFL=0.1$.}
% \label{fig:recons/ovs}
% \end{figure}

\subsection{High-resolution shock-capturing schemes}
As a classic issue in computational fluid dynamics, the central high-order discretized
schemes introduce high-frequency oscillations at the location of shocks and discontinuities. 
While the amplitude of these oscillations depends on the strength of the
shock among other factors, they seem to persist despite the size of the grid. This anomaly which is known as the Gibbs phenomena, can render the numerical scheme unacceptable. Different approaches have been developed to tackle this challenge. Among those, two fundamental methods have long and successfully been used in shock-capturing simulations: ENO and its successors such as WENO or TENO and TVD limiters \cite{pirozzoli2011numerical}.
Both schemes are robust in terms of capturing the discontinuities without spurious solutions, however the TVD schemes in general are known to be more dissipative \cite{arora1997well}.\\
In this work, we will make use of both methods for the following purposes: first, 
as any other numerical scheme, to be able to have accurate solutions in 
compressible shock-including simulation. Second, the PonD method depends on the
square root of local temperature at its core (see Eq. \eqref{eq:pond/discrete-vel}). 
Hence any oscillation that can push the temperature to negative values can not be allowed.
This is particularly concerning in high Mach flows \cite{zhang2011positivity}.

\subsubsection{WENO interpolation}
\begin{figure}[!t]
\centering
\includegraphics[width=0.7\textwidth]{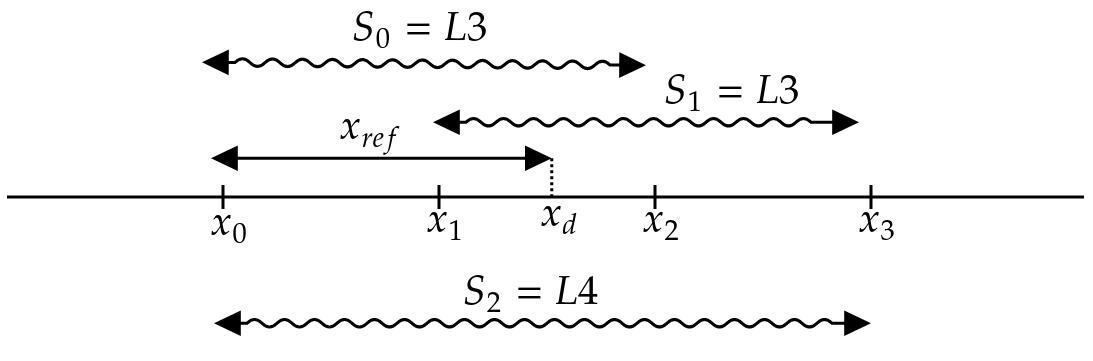}
\caption{Schematics of the fourth order WENO interpolation in one dimension.}
\label{fig:recons/weno-grid}
\end{figure}

\begin{figure}[!t]
    \centering
    \includegraphics[clip, trim= 0 0 0 2cm, width=0.7\textwidth]{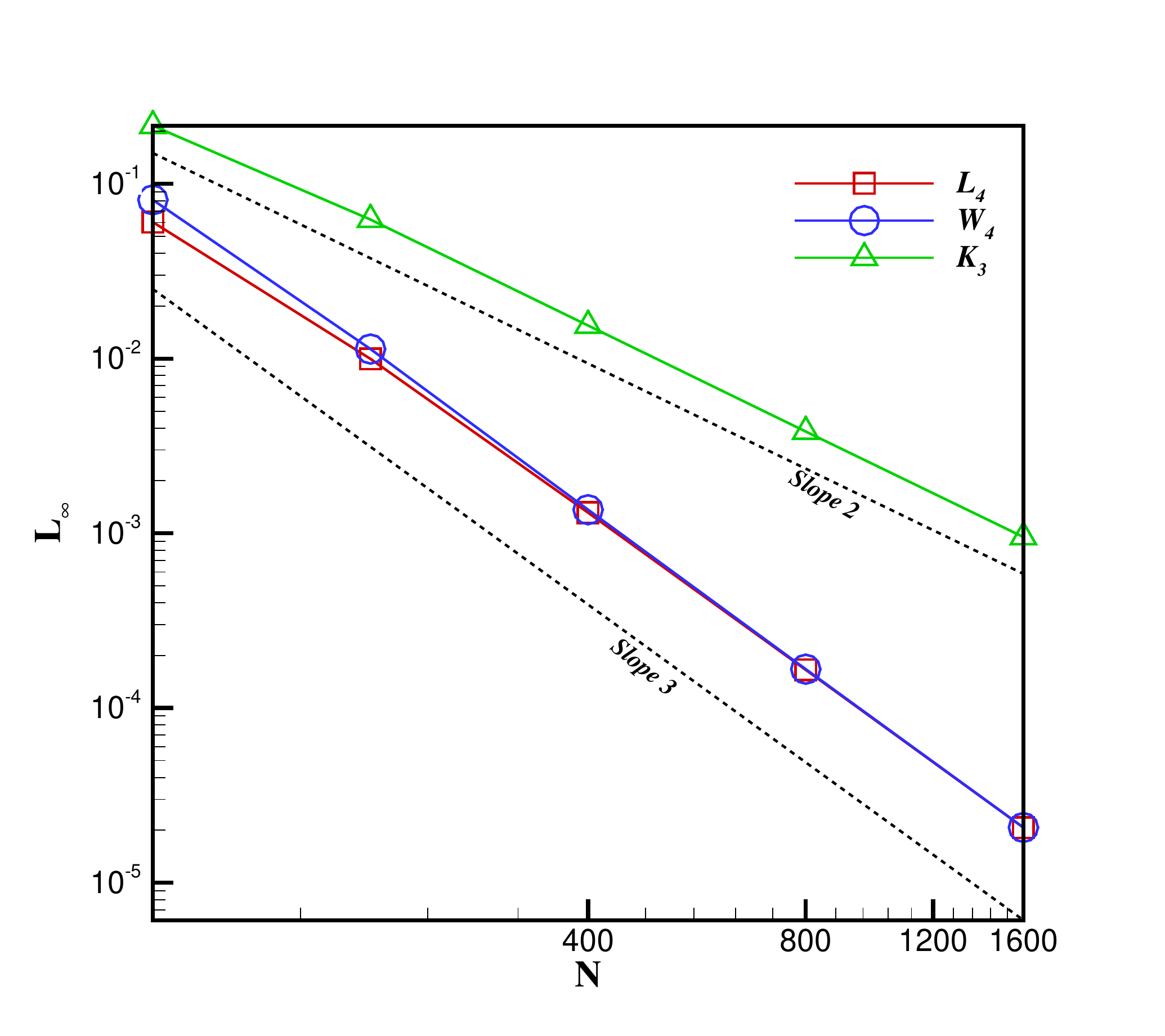}
\caption{OVS of $L_4$, $K_3$ and $W_4$ interpolation schemes. The results are obtained by solving the advection equation at $\cfl=0.1$.}
\label{fig:recons/ovs}
\end{figure}

While the main idea of WENO is based on choosing the smoothest kernel when 
interpolating at cell interfaces \cite{jiang1996efficient}, here we deal with
an interpolation problem at arbitrary points in space. However, one can use 
the very same concept in designing an oscillation-free interpolation function.
To construct a fourth-order interpolation scheme augmented with the essence of 
WENO, we consider a central 4-point kernel around the departure point. Figure \ref{fig:recons/weno-grid} illustrates a one-dimensional schematic of such setup. The main stencil is divided into two smaller stencils, i.e. $S_2 = S_1 \cup S_2$, where Lagrange polynomials are used in each of the stencils. 
The corresponding ideal weights of the sub-stencils are obtained as
\begin{align}
&\gamma_0 = 1-\frac{x_{ref}}{3\delta x},\\
&\gamma_1 = \frac{x_{ref}}{3\delta x},
\end{align} 
where $x_{ref} = x_d - x_0$ and  $x_d$ is the departure point.
If we are interested in interpolating the function $u(x)$, the final value of the interpolated function at the departure point becomes
\begin{align}
u(x_d) = \sum_{k=0}^{k=1} w_k u^{(k)},
\end{align}
where
\begin{align}
w_k = \frac{\tilde{w}_k}{\sum_k \tilde{w}_k}
\end{align}
is the normalized weights, $u^{(k)}$ is the interpolated value in each sub-stencil and
\begin{align}
\tilde{w}_k = \frac{\gamma_k}{(\epsilon+\beta_k)^2}.
\end{align}
The parameter $\epsilon$ is chosen as $10^{-6}$ to avoid zero denominator and $\beta_k$ is the smoothness indicator of each stencil defined as \cite{shu2020essentially}
\begin{align}
\beta_k = \sum_{l=1}^2 \int_{x_1}^{x_2} \frac{d^l}{dx^l}u^{(k)}(x),
\end{align}
where they are obtained as
\begin{align}
&\beta_0 = \nonumber\\
&\frac{1}{12}\left(13 u_0^2-52 u_0 u_1+26 u_0 u_2 + 64 u_1^2-76 u_1 u_2 + 25 u_2^2\right),\\
&\beta_1 = \nonumber\\
&\frac{1}{12}\left(25 u_1^2-76 u_1 u_2+26 u_1 u_3 + 64 u_2^2-52 u_2 u_3 + 13 u_3^2\right).
\end{align}
% The order of accuracy of the current scheme (which we entitle as $W_4$) is verified by solving the advection equation adopting the semi-Lagrangian method. The CFL number is  fixed at $0.1$ as before. Figure \ref{fig:recons/ovs-w4} shows that the  underlying order of accuracy of the $W_4$ scheme is recovered.
% \begin{figure}[!t]
% \centering
% \includegraphics[clip, trim = 0 0.5cm 0 1.5cm, width=\linewidth]{OVS-W4}
% \caption{OVS of the $W_4$ interpolation scheme. The results are obtained by solving the advection equation at $CFL=0.1$.}
% \label{fig:recons/ovs-w4}
% \end{figure}
Figure \eqref{fig:recons/ovs} shows the results of the order verification study (OVS) of the current scheme entitled as $W_4$ together with the $L_4$ and $K_3$ interpolation kernels. The results are the outcome of the semi-Lagrangian solution of the advection equation initialized with a Guassian profile. The CFL number is fixed at $0.1$. As expected, the underlying order of accuracy of all schemes are recovered.
\subsubsection{TVD Bspline limiters}
The so-called TVD scheme introduced by Harten, has been an effective tool 
in the class of high-resolution schemes to control the spurious oscillations \cite{arora1997well}. 
By definition, the total variation of a solution at time $n$ is
\begin{align}
{TV}[n] = \sum_i |u_{i+1}^n-u_i^n|
\end{align}
and a numerical scheme is said to be TVD if $ TV[n+1]\leqslant TV[n]$ \cite{jeng1995adaptive}.
Based on this definition, the so-called limiter functions are designed to retain
the smoothnes of the solutions at critical points. More detailed information
on limiters can be found in various researchs, such as \cite{jeng1995adaptive,shi1996fully,arora1997well}.\\

As discussed in section \ref{sec:recons/bspline}, Bsplines are smooth functions
and have better performance in interpolating fields with fluctuations than their 
counterparts due to their continuous derivatives \cite{koumoutsakos1997inviscid}. 
In our numerical experiments, we also observe the very good mass conserving property
of the $K_3$ kernel than $L_4$ and $W_4$. 
However, they still allow oscillations in discontinuous parts of the solution since 
negative weights are present (see Fig. \ref{fig:kernels}). Hence, to benefit
from the mass-conserving feature of the $K_3$ kernel in high Mach compressible
simulations, we aim at developing a TVD limiter function based on this kernel.\\

\begin{figure}[!t]
\centering
\includegraphics[width=0.7\textwidth]{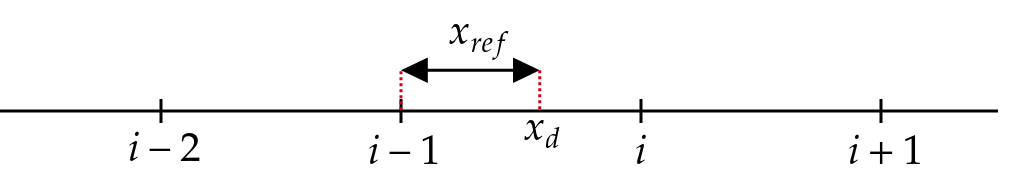}
\caption{Support points of the $K_3$ kernel around the interpolation point.}
\label{fig:recons/k3-grid}
In other words, we can rewrite the advection problem as
\begin{align}
u_i^{n+1} = \sum_{j=i-2}^{i+1} a_j u_j^n,
\end{align}
\end{figure}

\begin{figure}[!t]
\centering
\includegraphics[width=0.7\textwidth]{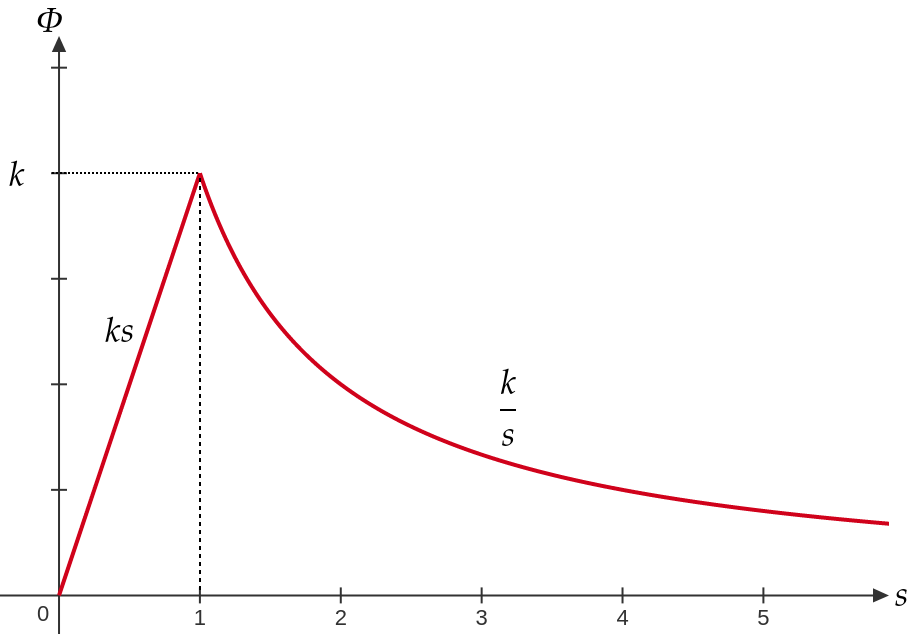}
\caption{Asymptotic form of the limiter function. For any point $s$, $\Phi(1/s) = \Phi(s)$.}
\label{fig:recons/TVD-k3}
\end{figure}

\begin{figure}[!t]
\centering
\includegraphics[width=0.7\textwidth]{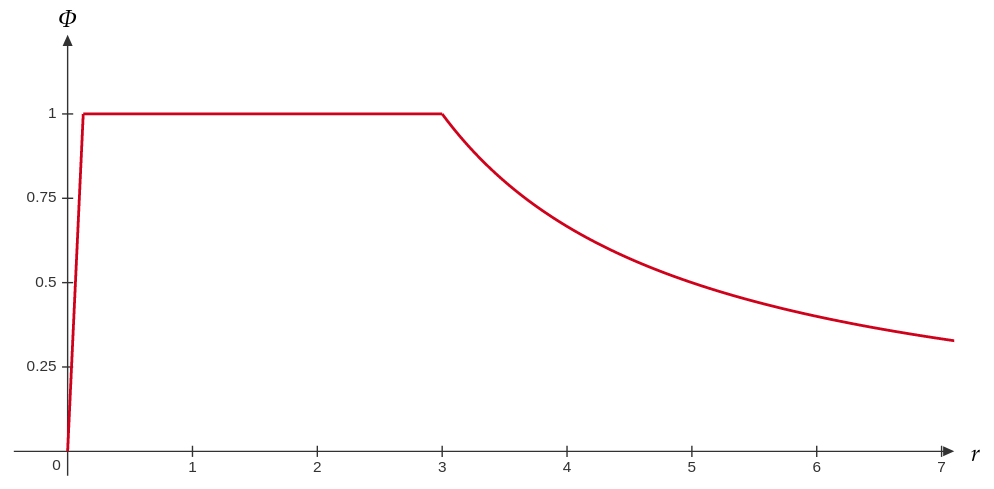}
\caption{The $K_3$ limiter function}
\label{fig:recons/TVD-limiter}
\end{figure}

According to Eq. \eqref{eq:recons/k3}, on a symmetric 4-point stencil, $K_3$ has the following weights
around the interpolation point (see Fig. \ref{fig:recons/k3-grid})
\begin{align}
a_{i-2}&= -\frac{1}{2}x_{ref}(1-x_{ref})^2,\nonumber\\
a_{i-1}&= -\frac{3}{2}(1-x_{ref})(x_{ref}-x_1)(x_{ref}-x_2),\nonumber\\
a_i    &= -\frac{3}{2}(x_{ref})(1-x_{ref}-x_1)(1-x_{ref}-x_2),\nonumber\\
a_{i+1}&= -\frac{1}{2}(1-x_{ref})x_{ref}^2,
\label{eq:recons/k3-weights}
\end{align}
where $x_{1,2} = (1\pm\sqrt{7})/3$ and $x_{ref}=x_d - x_{i-1}$. The weights \eqref{eq:recons/k3-weights} are
realized on a grid with $\delta x = 1$ and they sum to one. We can generalize this interpolation scheme
to a semi-Lagrangian advection problem with the CFL number defined as $\sigma =v\delta t/ \delta x$,
where $v$ is a characteristic velocity and $\delta t$ is the timestep. If we assume
that $|{\sigma}|<1$, then $x_d$ in Fig. \ref{fig:recons/k3-grid} is the departure point
found by traveling back in time from the monitoring point $x_i$ through the characteristic velocity
$v>0$.
where the weights $a_j$ are given by Eq. \eqref{eq:recons/k3-weights}.
However, it is logical to recast the weights as a function of the CFL number
and independent of the grid size. By doing so, we derive
\begin{align}
a_{i-2} &= -\frac{1}{2}\sigma^2(1-\sigma),\nonumber\\
a_{i-1} &= -\frac{1}{2}\sigma(3\sigma^2-4\sigma-1),\nonumber\\
a_i     &= -\frac{1}{2}(1-\sigma)(3\sigma^2-2\sigma-2),\nonumber\\
a_{i+1} &= -\frac{1}{2}\sigma(1-\sigma)^2,
\label{eq:recons/k3-weights-recast}
\end{align}
where $\sum_j a_j = 1$ and $\sigma>0$. Finally the advection problem with a $K_3$ interpolation scheme
can be rewritten as
\begin{align}
u_i^{n+1} = &u_i^n-\sigma\delta u_{i-1/2}^n - \frac{\sigma}{2}(1-\sigma)^2\delta u_{i+1/2}^n \nonumber\\
&+\frac{\sigma}{2}(1-\sigma)(1-2\sigma)\delta u_{i-1/2}^n + \frac{\sigma^2}{2}(1-\sigma)\delta u_{i-3/2}^n,
\end{align}
where $\delta u_{i\pm k/2} = u_{i\pm k/2 + 1/2}-u_{i\pm k/2 - 1/2}$ is the central difference operator.
We now proceed with the imposing the TVD limiter functions on the above equation
\begin{align}
u_i^{n+1} &= u_i^n-\sigma\delta u_{i-1/2}^n - \frac{\sigma}{2}(1-\sigma)^2\delta u_{i+1/2}^n \Phi_{i+1/2} \nonumber\\
&+\frac{\sigma}{2}(1-\sigma)(1-2\sigma)\delta u_{i-1/2}^n \Phi_{i-1/2}\nonumber\\
&+ \frac{\sigma^2}{2}(1-\sigma)\delta u_{i-3/2}^n \Phi_{i-3/2}.
\label{eq:recons/k3-expanded}
\end{align}
To relate the limiters to the oscillations in the solution, they are commonly defined as a 
function of slope ratios 
\begin{align}
\Phi_{i+1/2}=\Phi(r_{i+1/2}^+),
\end{align} 
where the superscript denotes the positive CFL number and 
\begin{align}
r_{i+1/2}^+=\frac{\delta u_{i-1/2}}{\delta u_{i+1/2}}=\frac{u_i-u_{i-1}}{u_{i+1}-u_i},
\end{align}
is an example of slope ratios which in general is defined as the upwind difference devided 
by the central difference \cite{arora1997well}.
One can write the following 
\begin{align}
&\Phi_{i+1/2} \delta u_{i+1/2} = \Phi_{i+1/2} \frac{\delta u_{i+1/2}}{\delta u_{i-1/2}}\delta u_{i-1/2} = 
\frac{\Phi(r_{i+1/2}^+)}{r_{i+1/2}^+}\delta u_{i-1/2},\\
&\Phi_{i-3/2} \delta u_{i-3/2} = \Phi_{i-3/2} \frac{\delta u_{i-3/2}}{\delta u_{i-1/2}}\delta u_{i-1/2} = 
\frac{\Phi(r_{i-3/2}^-)}{r_{i-3/2}^-}\delta u_{i-1/2}.
\end{align} 
To keep consistency, we rewrite the latter as
\begin{align}
\frac{\Phi(r_{i-3/2}^-)}{r_{i-3/2}^-}\delta u_{i-1/2} = \Phi\left(\frac{1}{r_{i-1/2}^+}\right) {r_{i-1/2}^+}\delta u_{i-1/2}.
\end{align} 
Adopting the notations $r=r_{i+1/2}^+$ and $s=r_{i-1/2}^+$, we can rewrite Eq. \eqref{eq:recons/k3-expanded} in the following compact form
\begin{align}
&u_i^{n+1} = u_i^n-\delta u_{i-1/2}^n\times\nonumber\\
&\sigma \left(1+\frac{1}{2}(1-\sigma)\left[(1-\sigma)\frac{\Phi(r)}{r} + (2\sigma-1)\Phi(s)-\sigma s \Phi(\frac{1}{s})\right] \right).
\end{align}
Finally, repeating the same procedure for the negative CFL number,
the necessary condition for the scheme to be TVD is found as
\begin{align}
-\frac{2}{1-|\sigma|}\leqslant (1-|\sigma|)\frac{\Phi(r)}{r} + (2|\sigma|-1)\Phi(s)-|\sigma| s \Phi\left(\frac{1}{s}\right)\leqslant \frac{2}{|\sigma|}.
\label{eq:recons/TVD-condition}
\end{align}
To find the suitable range of the limiter function, the following conditions are considered
\begin{align}
\Phi(r) = 0, \ \text{if} \ r\leqslant 0, \nonumber\\
\Phi(r) > 0, \ \text{if} \ r> 0, \nonumber\\
\Phi(r) = 1, \ \text{if} \ r=1.
\label{eq:recons/TVD-constraint}
\end{align}
Two distinct steps are considered. First we assume $r>0, s<0$. Considering the
constraints \eqref{eq:recons/TVD-constraint}, Eq. \eqref{eq:recons/TVD-condition} becomes
\begin{align}
\Phi(r) \leqslant \frac{2r}{|\sigma|(1-|\sigma|)},
\end{align}
where the most stringent condition gives
\begin{align}
\Phi(r) \leqslant 8r.
\label{eq:recons/TVD-con-1}
\end{align}
In the second step, we assume the opposite as before, i.e. $r<0, s>0$ which renders
Eq. \eqref{eq:recons/TVD-condition} as
\begin{align}
-\frac{2}{1-|\sigma|}\leqslant  (2|\sigma|-1)\Phi(s)-|\sigma| s \Phi\left(\frac{1}{s}\right)\leqslant \frac{2}{|\sigma|}.
\end{align}
To have a sensible evaluation of the function $\Phi(1/s)$, we assume that the limiter function has the form $\Phi(s) =
\text{min}(ks,k/s)$ as shown in Fig. \ref{fig:recons/TVD-k3}. This form of definition implies $\Phi(1/s) = \Phi(s)$.
 Hence, we can write the above equation as
\begin{align}
-\frac{2}{1-|\sigma|}\leqslant \left( (2|\sigma|-1) - s|\sigma|\right)\Phi(s) \leqslant \frac{2}{|\sigma|}.
\end{align}
The following restraints can be derived
\begin{align}
&\Phi(s) \leqslant \frac{2}{(1-|\sigma|)\left(s|\sigma|-2|\sigma|+1\right)}, \ |\sigma|<0.5 , \nonumber\\
&\Phi(s) \leqslant \frac{2}{\eta\mid s|\sigma|-2|\sigma|+1\mid}, \ |\sigma|>0.5 ,
\label{eq:recons/TVD-con-2}
\end{align}
where
\begin{align}
\eta = \Bigg\{
\begin{tabular}{cc}
$|\sigma|$    ,&  $0<s<\frac{2|\sigma|-1}{\mid\sigma}$,\\
$1-|\sigma|$  ,& $s>\frac{2|\sigma|-1}{|\sigma|}$.
\end{tabular}
\end{align}
Finally, collecting the conditions \eqref{eq:recons/TVD-con-1} and \eqref{eq:recons/TVD-con-2}, the most stringent limiter 
is obtained as
\begin{align}
\Phi(r) =\max\left[0, \min\left(8r,1,\frac{2}{r-1}\right)\right],
\label{eq:recons/TVD-limiter}
\end{align}
which is illustrated in Fig. \ref{fig:recons/TVD-limiter}.

\begin{figure}[!t]
\centering
\includegraphics[clip, trim= 0 0 0 2cm,width=0.7\textwidth]{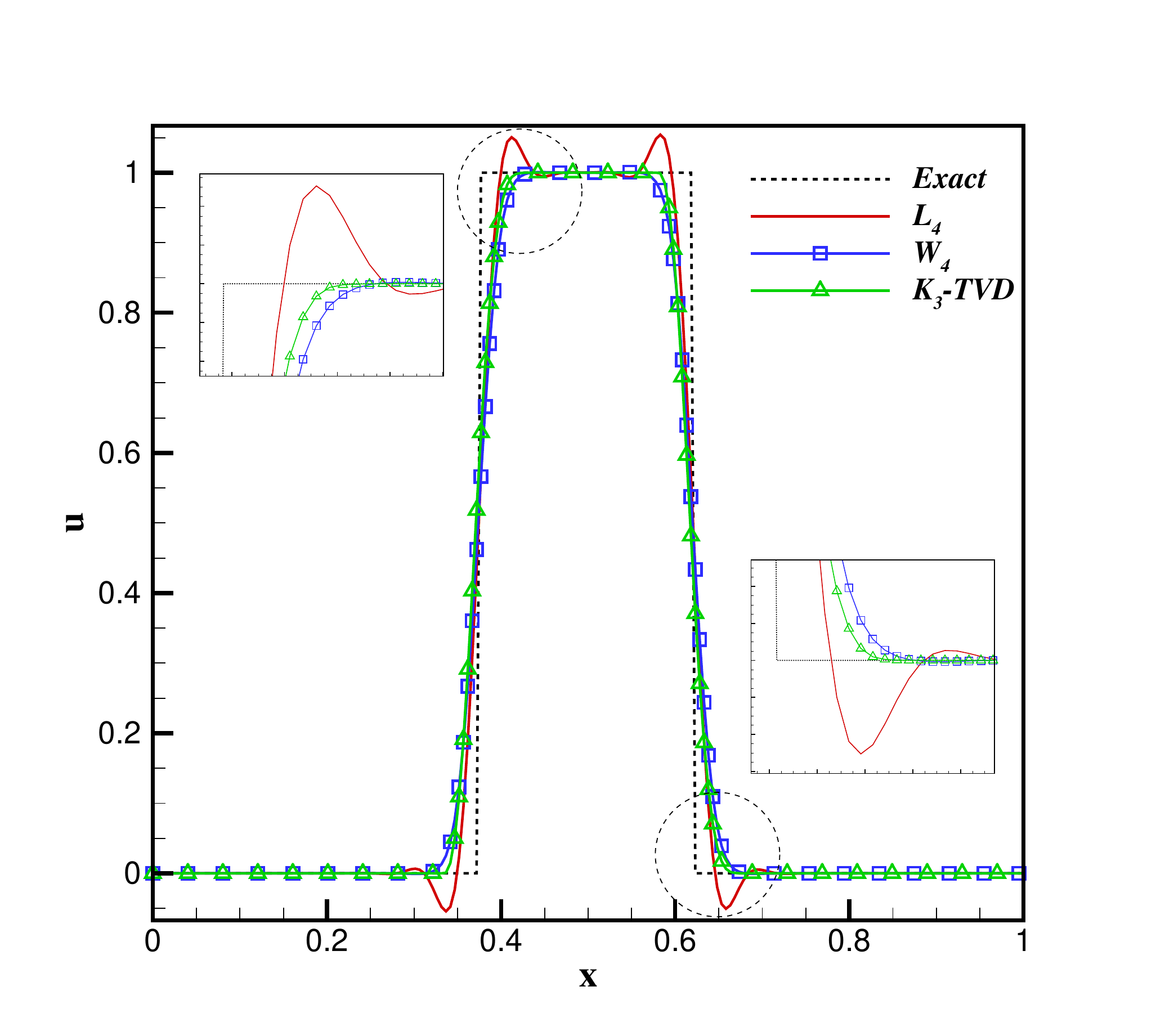}
\caption{Semi-Lagrangian solution of the advection equation with $L_4$, $W_4$ and $K_3-\tvd$ schemes after two periods.}
\label{fig:recons/advection}
\end{figure}

To put the developed schemes into test, we solve the linear advection equation
\begin{align}
&u_t + au_x = 0, \ 0\leqslant x \leqslant 1, \nonumber \\
&u(x,0) = \Bigg\{ 
\begin{tabular}{cc}
$1$ , & $3/8 \leqslant x \leqslant 5/8$,\\
$0$ , & otherwise,
\end{tabular}
\end{align}
where $a$ is the constant speed. We compute the solution up to $t=2$ which amounts to
two total periods. A semi-Lagrangian scheme is adopted. 
$200$ points are used to discretize the space and the CFL number is
$\sigma = a\delta t / \delta x = 0.4$.
Figure \ref{fig:recons/advection} shows the results for the $L_4$, $W_4$ and $K_3-\tvd$ schemes after two periods. As we can see, 
the oscillations at the discontinuities are successfully eliminated using the TVD and WENO schemes.
.

`
\section{Spectral Analysis}
Since the present model is based on semi-Lagrangian advection, first we discuss the spectral properties of a linear interpolation scheme such as the dissipation and dispersion errors against the reduced wave number $\zeta$ in the interval $[0,\pi]$. Considering the filed $u(x,t)$ that is advected by a constant velocity $a$ and assuming a general discrete form $u_i(t)=\hat{u}(t)\exp(\euler i\zeta)$, the solution via the semi-Lagrangian advection in the time-interval of $[0,t]$ is obtained as
\begin{align}
    u_i(t)=\hat{u}(0){\rm e}^{\euler i\zeta}\sum_{j=-l}^r w_j{\rm e}^{\euler j\zeta},
\end{align}
where $w=w(\sigma)$ is the interpolation weights and $\sigma = at/\delta x$ is the \cfl  number.
Finally, the modified wave number is derived with the following real and imaginary parts
\begin{align}
    \mathfrak{Re} (\Psi) &= \frac{1}{\sigma}\ln R\nonumber,\\
    \mathfrak{Im} (\Psi) &= -\frac{\theta}{\sigma},
    \label{eq:spectral-theory}
\end{align}
where $(R,\theta)$ are the modulus and argument of the complex term $\sum w_j \exp(\euler j\zeta)$, respectively. With this, one can write the discrete solution as \cite{pirozzoli2006spectral}
\begin{align}
    u_i(t) = \hat{u}(0){\rm e}^{\euler i\zeta}{\rm e}^{-\euler\sigma\Psi},
\end{align}
where spectral schemes feature $\Psi(\zeta)=\zeta$.

\begin{figure}[!t]
    \centering
    \includegraphics[clip, trim= 0 0.5cm 0 1cm,width=0.7\textwidth]{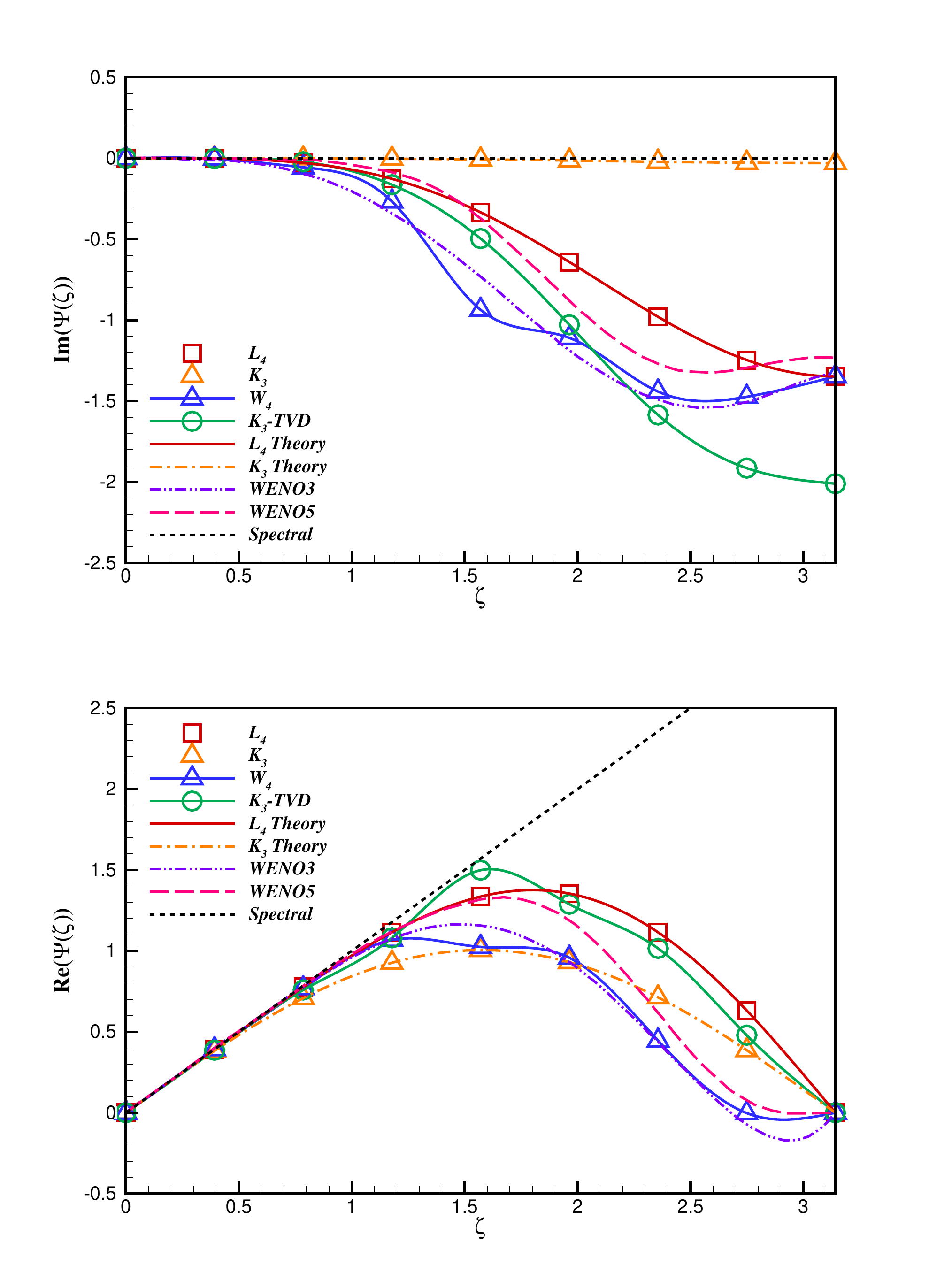}
    \caption{Approximate dispersion relation for various linear and shock-capturing schemes: dissipation (top) and dispersion (bottom).}
    \label{fig:ADR}
\end{figure}

While this theory can be employed for analysis of the dissipation and dispersion errors of linear models, the spectral properties of nonlinear schemes such as the shock-capturing methods are derived using the approximate dispersion relation (ADR) proposed by Pirozzoli \cite{pirozzoli2006spectral}. Assuming a sinusoidal initial condition with different reduced wave numbers, we apply the ADR method to the WENO and $K_3-\tvd$ schemes.
Figure \ref{fig:ADR} shows the spectral properties of both schemes along with the dispersion and dissipation of the linear $L_4$ and $K_3$ functions. As we see, the results from the ADR analysis coincide with those obtained from theory (Eq. \eqref{eq:spectral-theory}) for the linear schemes. It is also observed that while the $L_4$ scheme has a superior dispersion property, the $K_3$ scheme features much less dissipation almost as the spectral. Regarding the non-linear schemes, the $K_3- \tvd$ has an improved dispersion property and follows the spectral closely up to the reduced wave number $\zeta=\pi/2$, while becoming more dissipative. On the other hand, the wave-resolution property of the WENO-interpolation scheme follows the spectral up to $\zeta=1.2$, i.e. low to moderate wave numbers. Finally, all schemes are stable through the entire range of wave numbers, where ${\rm Im}(\Psi)\leqslant 0$.

\section{Benchmarks}

In this section, we will use $W_4$ and $K_3-\tvd$ schemes in the PonD framework. To verify the accuracy and robustness, a various number of standard compressible benchmarks are considered.
In each test case, we will assess the performance of both lattice geometries.
In the following, we will first consider the two most common mild shock-tube 
problems, the Sod and the Lax shock-tube. Then the simulation of the 
Shu-Osher problem as a rather stronger case will be presented followed 
by considering some strong cases.
In our TVD scheme, we choose the zeroth moment of each set of populations as the slope ratio determinant for the corresponding population, i.e. density for the $f$ population and total energy for the $g$, unless stated otherwise. The CFL number in our simulations are based on the maximum magnitude of the discrete velocities
\begin{align}
\cfl =\frac{\max(\|\vi \|)\delta t}{\delta x}
\end{align}
and is fixed to $\cfl=0.2$. The viscosity is chosen small enough such that the simulations are stable ($\nu\delta t/\delta x^2\approx\mathcal{O}(10^{-6}-  10^{-4}$). We use the standard $D2Q9$ lattice in all simulations.
%****************************************************
\subsection{Sod problem}
\begin{figure}
\centering
\includegraphics[clip, trim = 0 0 0 1cm, width=\textwidth]{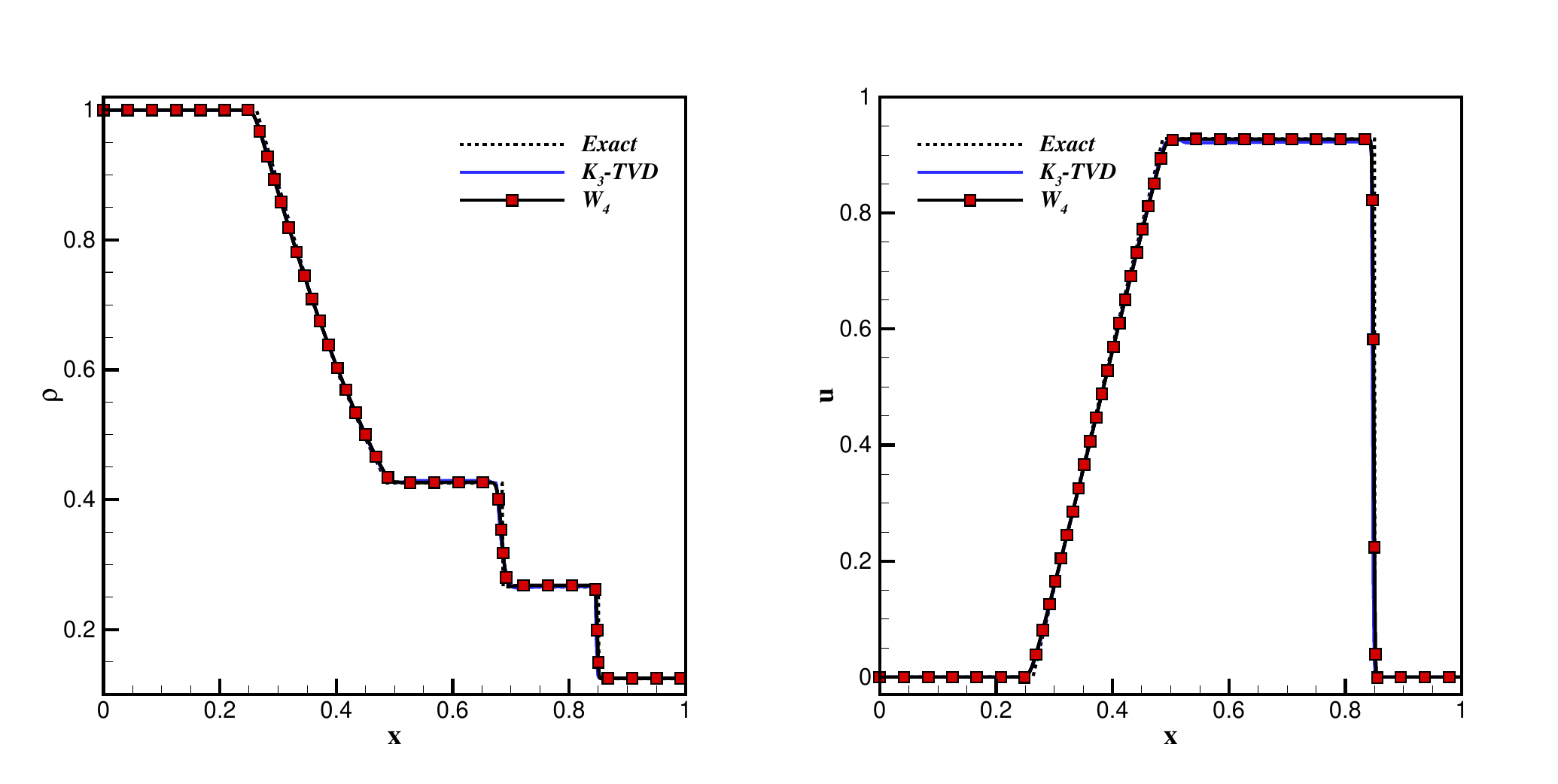}
\caption{Sod shock-tube problem: density distribution (left) and velocity distribution (right).
"Exact" represents the solution of the Riemann problem.
$Nx=600$ points are used to discretize the domain.}
\label{fig:compressible/sod-9}
\end{figure}

The initial condition for the Sod problem is \cite{SOD19781}:
\begin{align}
(\rho,u,p)=\left\{
  \begin{tabular}{lr}
  $(1,0,1)$  & if $0\leqslant x < 0.5$, \\
    $(0.125,0,0.1)$ & if $0.5\leqslant x \leqslant 1$,
\end{tabular}
\right.
\end{align}
where $x$ is the non-dimensional length of the tube and the final simulation time is $t=0.2$.
% The parameters of the simulation is listed in table \ref{table:compressible/sod}. 
$Nx=600$ points are used to discretize the domain.
Figure \ref{fig:compressible/sod-9} shows the results for the density and velocity distributions using the $K_3-\tvd$ and $W_4$ schemes. Both schemes have similar performances, where the results agree well with the exact solution. It is also  visible that both schemes have successfully captured the shock and the discontinuity free of oscillations.

%****************************************************
\subsection{Lax problem}
\begin{figure}
\centering
\includegraphics[clip, trim = 0 0 0 1cm, width=\textwidth]{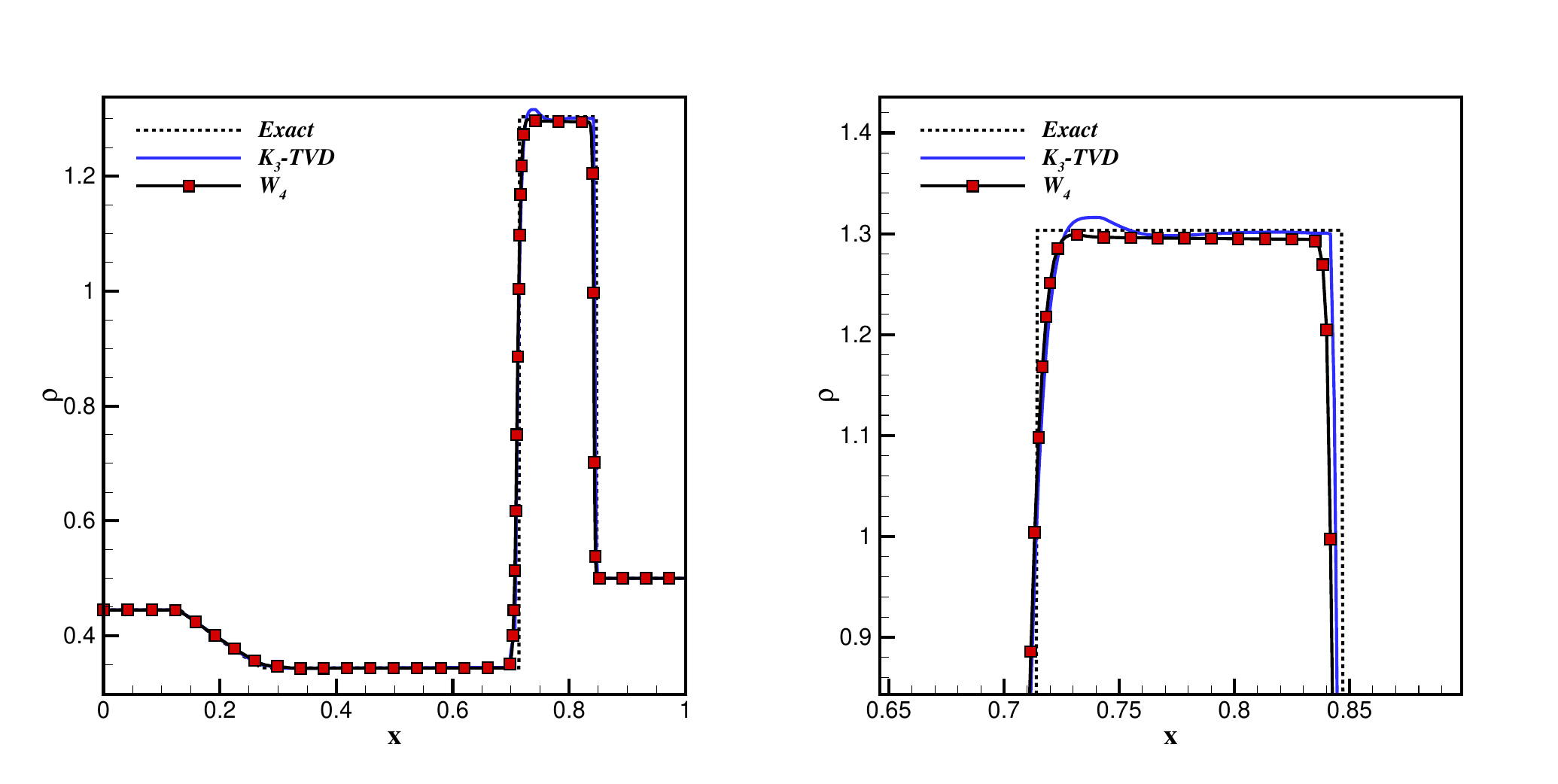}
\caption{Lax shock-tube problem: density distribution for $K_3-\tvd$ and $W_4$ schemes (left) and a zoom of the density distribution around the peak values (right).
"Exact" represents the solution of the Riemann problem. $Nx=600$ points are used to discretize the domain. }
\label{fig:compressible/lax-k3-w4}
\end{figure}

We consider the Lax problem \cite{lax1954weak} with the initial condition of 
\begin{align}
(\rho,u,p)=\left\{
  \begin{tabular}{lr}
  $(0.445,0.689,3.528)$  & if $0\leqslant x < 0.5$, \\
    $(0.5,0,0.571)$ & if $0.5\leqslant x \leqslant 1$.
\end{tabular}
\right.
\end{align}
The final simulation time is $t=0.14$.
The domain is discretized using $Nx=600$ points. Figure \ref{fig:compressible/lax-k3-w4}
shows the results for the density  distribution. As we can observe, the monotonicity of the 
solution is well preserved near the contact discontinuity and the shock-wave in both schemes.
Furthermore, the results are in good agreement with the exact solution.
%****************************************************
\subsection{Shock density-wave interaction}
\begin{figure}[!t]
\centering
\includegraphics[clip, trim = 0 0 0 1cm, width=\textwidth]{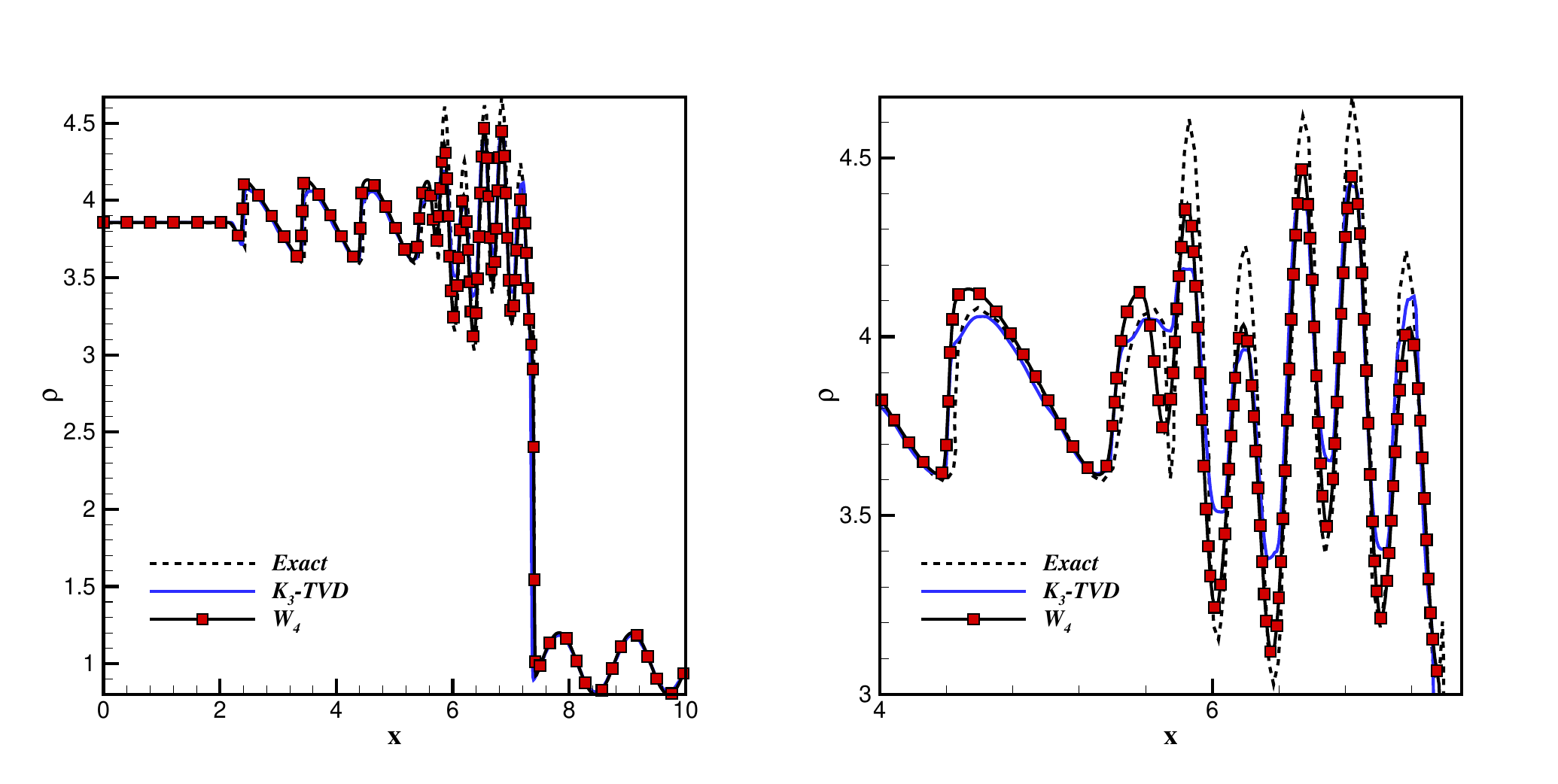}
\caption{Shu-Osher problem using the $K_3-\tvd$ and $W_4$ schemes: density distribution (left) and a zoom of the density distribution (right). $Nx=800$ points are used.}
\label{fig:compressible/shu-osher-9}
\end{figure}

\begin{figure}[!t]
\centering
\includegraphics[clip, trim = 0 0 0 1cm, width=\textwidth]{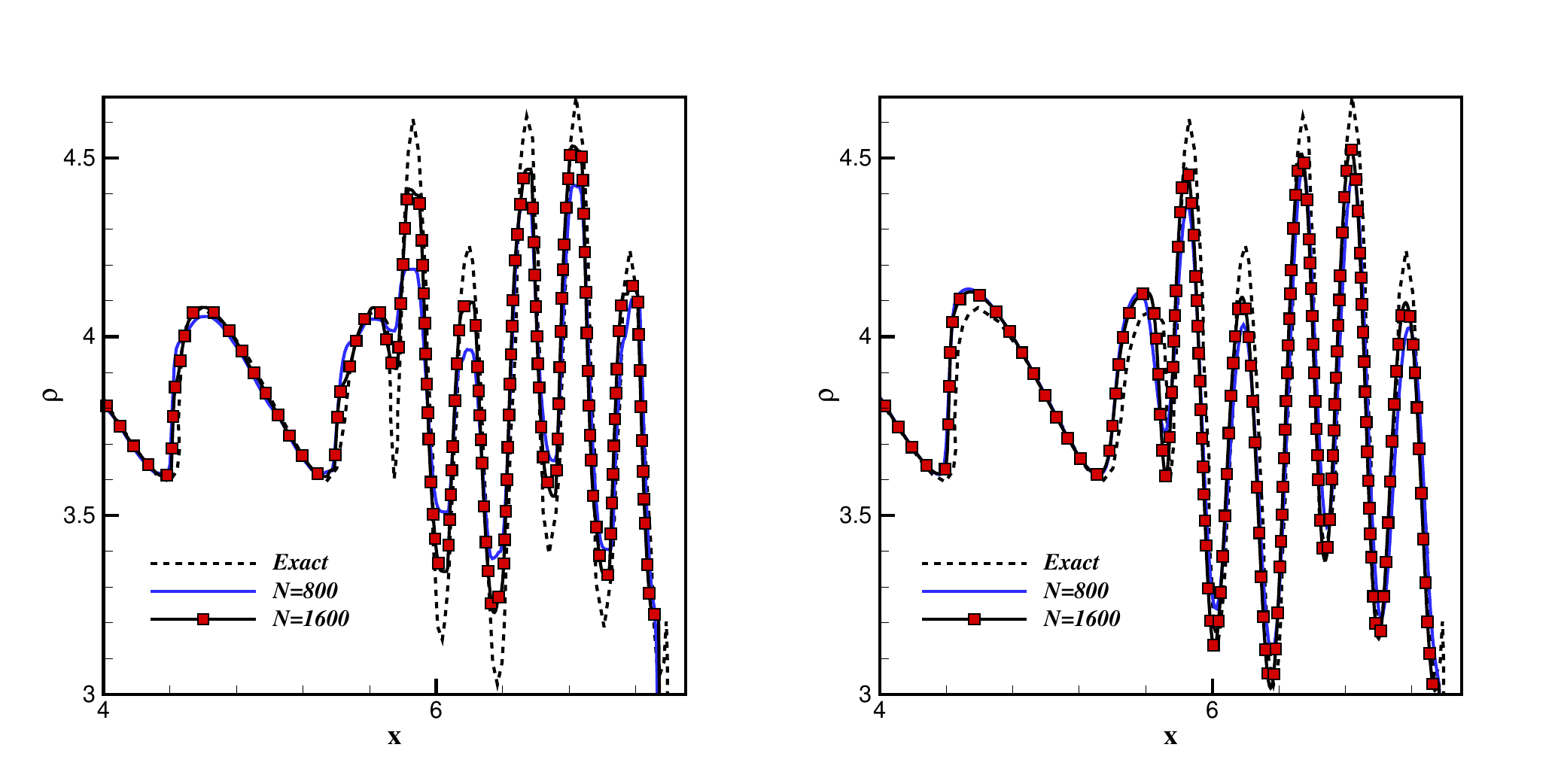}
\caption{Shu-Osher problem: A zoom of the density distribution using the $K_3-\tvd$ scheme(left) and $W_4$ scheme (right) at different resolutions.}
\label{fig:compressible/shu-osher-9-high-res}
\end{figure}

Also known as the Shu-Osher problem \cite{SHU198932}, a Mach 3 shock wave interacts with a 
perturbed density field leading to discontinuities and small structures. The initial condition
for this problem is
\begin{align}
(\rho,u,p)=\left\{
  \begin{tabular}{lr}
  $(3.857,2.629,10.333)$  & if $0\leqslant x < 1$, \\
    $(1+0.2\sin(5x),0,1)$ & if $1\leqslant x \leqslant 10$.
\end{tabular}
\right.
\end{align}
The final simulation time is $t=1.8$.
% Figure \ref{fig:compressible/shu-osher-25} gives a comparison of the computed density distribution
% for both schemes using the D2Q25 lattice at $\cfl=0.1$ and $\nu=10^{-3}$. It is clear that the location of the shock does not coincide with the exact solution. Further decreasing the viscosity does not effect the location of the shock, however more dissipation was observed as expected. Moreover, the WENO scheme underestimates the amplitude of the post-shock waves while the TVD scheme has a better performance. Furthermore, in the case of the TVD scheme, we can see that the location of the shock is improved by decreasing the CFL number. Finally, increasing the resolution results in converging to the exact solution. 
% At this stage, it can be noticed that the post-shock density waves are properly captured  and the 
% shock-wave is captured free of oscillations.
% \begin{figure*}
% \centering
% \includegraphics[clip, trim = 0 0 0 1cm, width=\textwidth]{Shu-Osher-25}
% \caption{Shu-Osher problem: density distribution (D2Q25). Comparison of WENO and TVD schemes at $\cfl=0.1$, $\nu=10^{-3}$ 
%  (left) and $K_3-\tvd$ scheme at $\cfl=0.02$, $\nu=10^{-4}$ (right). 800 grid points were used.}
% \label{fig:compressible/shu-osher-25}
% \end{figure*}
$Nx=800$ points are used for discretizing the space.
Figure \ref{fig:compressible/shu-osher-9} shows the results for the density distribution compared to the exact solution.
We observe that the WENO scheme acts inferior in this simulation. As seen in Fig. \ref{fig:compressible/shu-osher-9-high-res}, the acoustic waves are overestimated and they show a convergent behavior, i.e. they do not improve by increasing the resolution. On the other hand, the TVD scheme captures the proper amplitude of the acoustic waves and the entropy waves are better resolved with increasing the resolution. Nevertheless, the shock is captured free of oscillations in both schemes.

% \begin{figure*}[!t]
% \centering
% \includegraphics[clip, trim = 0 0 0 1cm, width=\textwidth]{Shu-V25-Reg-800-Final}
% \caption{Shu-Osher problem using the $K_3-\tvd$ and $W_4$ schemes with the $D2Q25$ lattice: density distribution (left) and a zoom of the density distribution (right). CFL=0.2 and $N=800$ points are used.}
% \label{fig:compressible/shu-osher-25-reg}
% \end{figure*}
%****************************************************
\subsection{Strong shock-tube}
\begin{figure}
\centering
\includegraphics[clip, trim = 0 0 0 1cm, width=\textwidth]{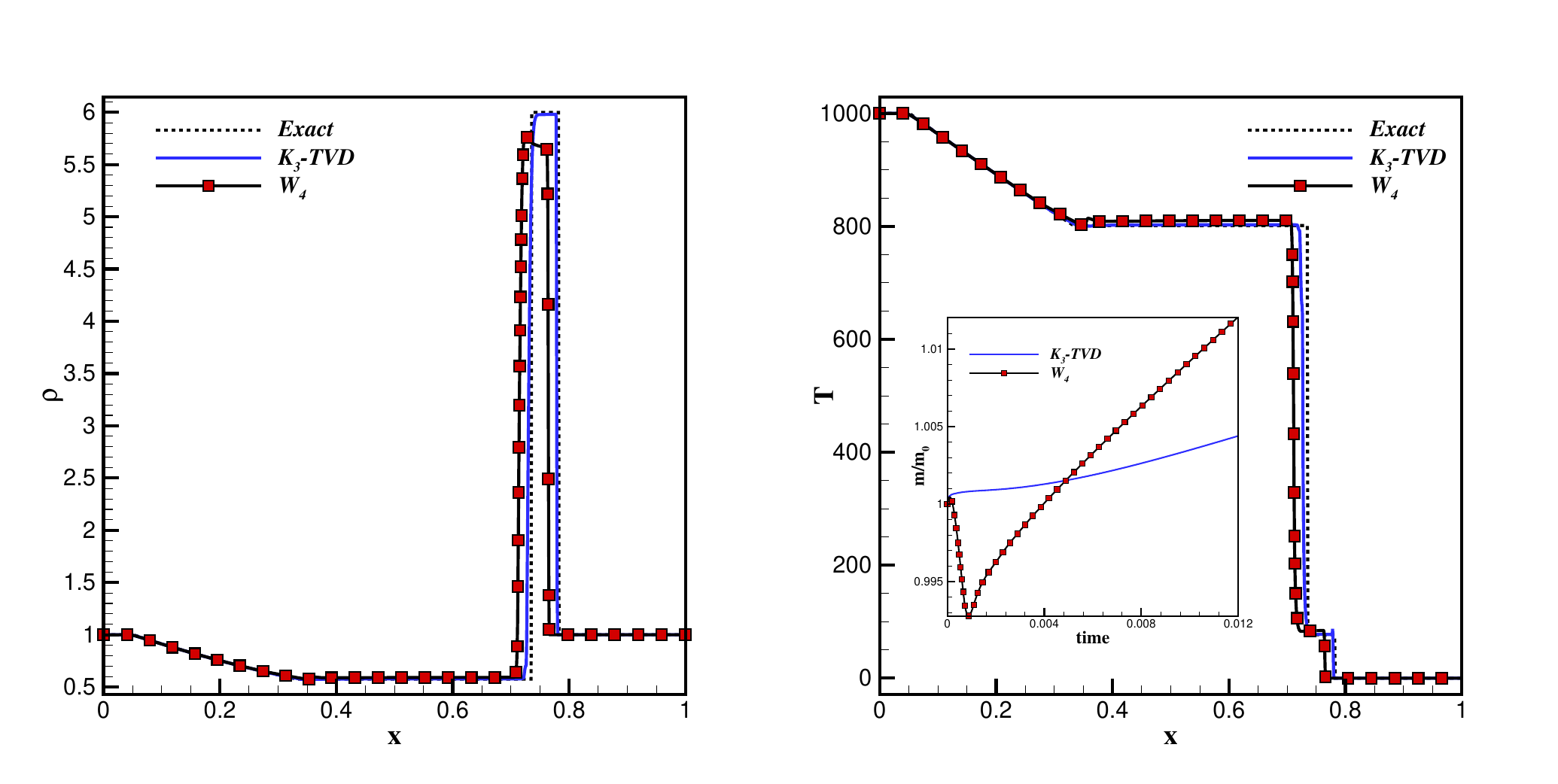}
\caption{Strong shock-tube problem: density distribution (left) and temperature profile (right) using the $K_3-\tvd$ and $W_4$ schemes with $Nx=1600$ grid points. "Exact"
represents the solution of the Riemann problem. The inset at the right picture shows the dimensionless mass of the domain for both schemes throughout the simulation.}
\label{fig:compressible/toro-v9}
\end{figure}

We consider a strong shock-tube case where the value of the Mach number reaches to 198 \cite{toro2012flux}.
The initial conditions are
\begin{align}
(\rho,u,p)=\left\{
  \begin{tabular}{lr}
  $(1,0,1000)$  & if $0\leqslant x < 0.5$, \\
    $(1,0,0.01)$ & if $0.5\leqslant x \leqslant 1$,
\end{tabular}
\right.
\end{align}
where the temperature ratio of both sides is $10^5$. The final simulation time is $t=0.012$.
Figure \ref{fig:compressible/toro-v9}
shows a comparison of the computed density field against the exact solution, using $Nx=1600$ grid points. As we can notice,
the location of the shock and the contact discontinuity is captured free of oscillations by both schemes,
while the WENO scheme does not coincide with the exact solution. This is visible in both density and temperature profiles. This could be explained
by looking at the evolution of total mass of the domain throughout the simulation. The TVD scheme features a very well mass conservation 
than that of the WENO (see inset). Eventually, the mass change in the WENO scheme leads to deviations in capturing the correct location of the shock front.
\subsection{Le Blanc Problem}
\begin{figure}[!t]
     \centering
     \includegraphics[clip, trim = 0 0 0 1cm, scale=0.65]{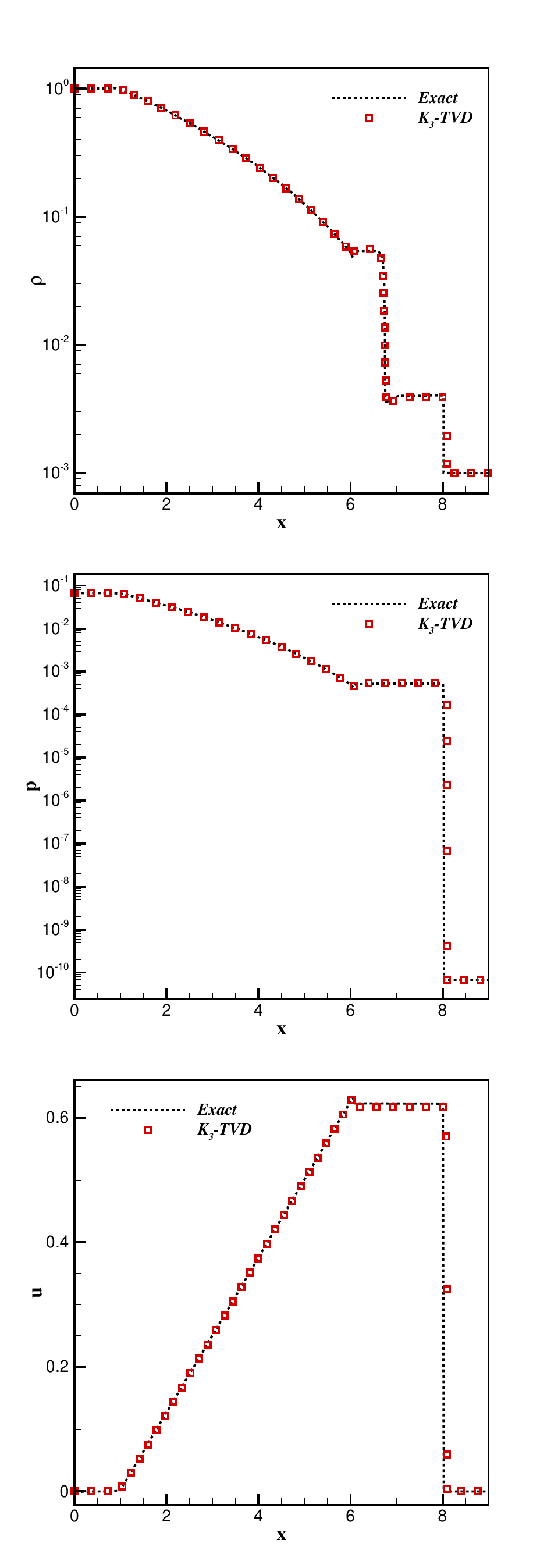}
     \caption{Le Blanc problem: from top to bottom; density, pressure and velocity distribution using the $K_3$ TVD scheme. $Nx=4000$ points are used.}
     \label{fig:leblanc}
 \end{figure}

Known as an extreme test case with very strong discontinuities, the Le Blanc problem \cite{LOUBERE2005105} has the following initial conditions
\begin{align}
(\rho,u,p)=\left\{
  \begin{tabular}{lr}
  $(1,0,2/3 \times 10^{-1})$  & if $0\leqslant x < 3$, \\
    $(10^{-3},0,2/3 \times 10^{-10})$ & if $3\leqslant x \leqslant 9$.
\end{tabular}
\right.
\end{align}
The final simulation time is $t=6$ and $Nx=4000$ grid points are used. The adiabatic coefficient is set to $\gamma=5/3$.
Figure \ref{fig:leblanc} shows the results for the density, pressure and velocity distribution using the $K_3$-TVD scheme, where
they agree well with the reference solutions. The WENO scheme failed this simulation.

%****************************************************
\subsection{Double Mach Reflection}
\begin{figure}[!t]
\centering
\includegraphics[clip, trim = 0 1.5cm 0.2cm 1cm, width=\textwidth]{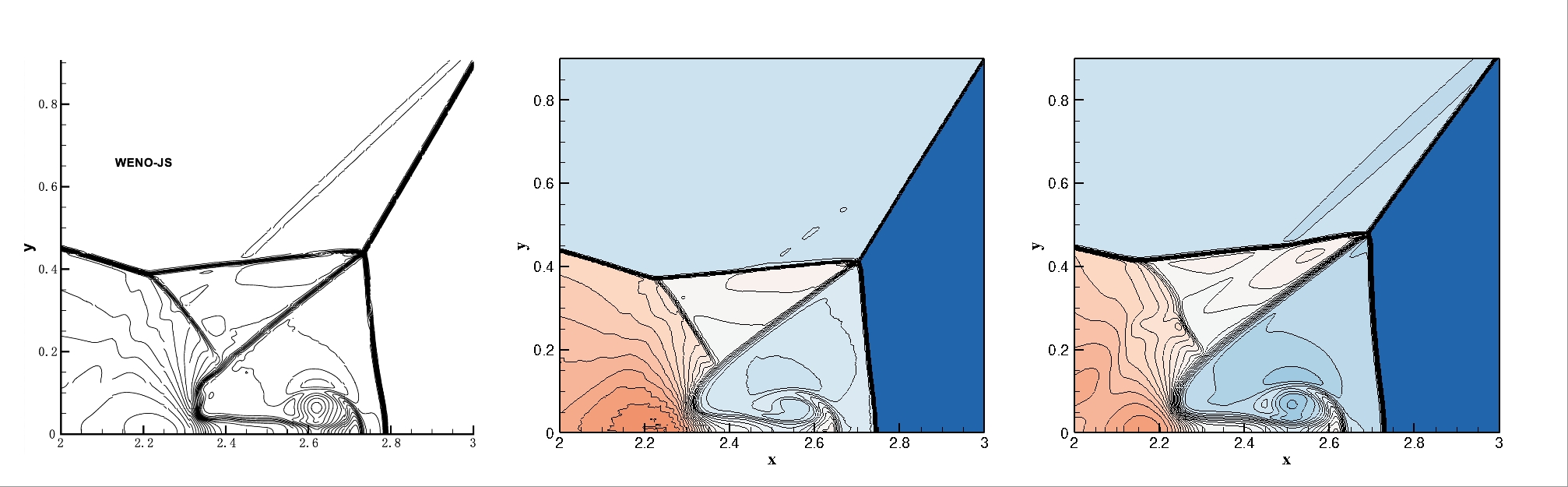}
\caption{Double Mach reflection of a strong shock: density contours in reference solution using WENO-JS-5 \cite{FU2016333} (left), present model using the $K_3$-TVD scheme (middle) and the $W_4$ scheme (right). 43 contours are drown between 1.887 and 20.9. Reference figures reprinted from \cite{FU2016333} with permission from Elsevier.}
\label{fig:DMR-V9}
\end{figure}

The initial condition for this case is \cite{WOODWARD1984115}
\begin{align}
&(\rho,u,v,p)\nonumber\\
&=\left\{
  \begin{tabular}{ll}
  (1.4,0,0,1),  & if $y>1.732(x-0.1667)$, \\
    (8,7.145,-4.125,116.8333), & otherwise,
\end{tabular}
\right.
\end{align}
which describes a right-moving Mach 10 incident shock wave initially placed at x=0.1667, with an incidence angle of $60^{\circ}$ to the $x$-axis. The computational domain is [0,4]$\times$[0,1] and the final simulation time is $t=0.2$. The post-shock condition is applied to the left boundary, whereas zero-gradient of all fluid variables is applied to the right boundary. At the bottom boundary, the post-shock condition is imposed from $x=0$ to $x=0.1667$, while a reflecting wall condition is enforced from $x=0.1667$ to $x=4$. The top boundary is treated such that all the fluid variables follow the evolution of the traveling shock-wave. For this simulation, we choose $Nx\times Ny=1201\times 301$ grid points.
Figure \ref{fig:DMR-V9} shows the results for both 
$K_3$-TVD and $W_4$ schemes compared to the reference solution \cite{FU2016333}, where temperature is used as the slope ratio determinant for the TVD scheme. To facilitate the comparison, only a part of the domain is presented. We observe that with the TVD scheme, the flow features and their locations are in good match with the reference solution. On the other hand, a close assessment of the results corresponding to the WENO scheme reveals some discrepancies with respect to the reference solution, such as the location of the second triple point. However, less dissipation than the TVD scheme is apparent in the jet area, where more flow structures are resolved. Overall, the comparison of density contour lines suggests that the TVD scheme is more accurate in this simulation.\\

\subsection{Astrophysical jet}
\begin{figure}
\centering
\includegraphics[clip, trim = 0 0.1cm 0.1cm 1cm, width=\textwidth]{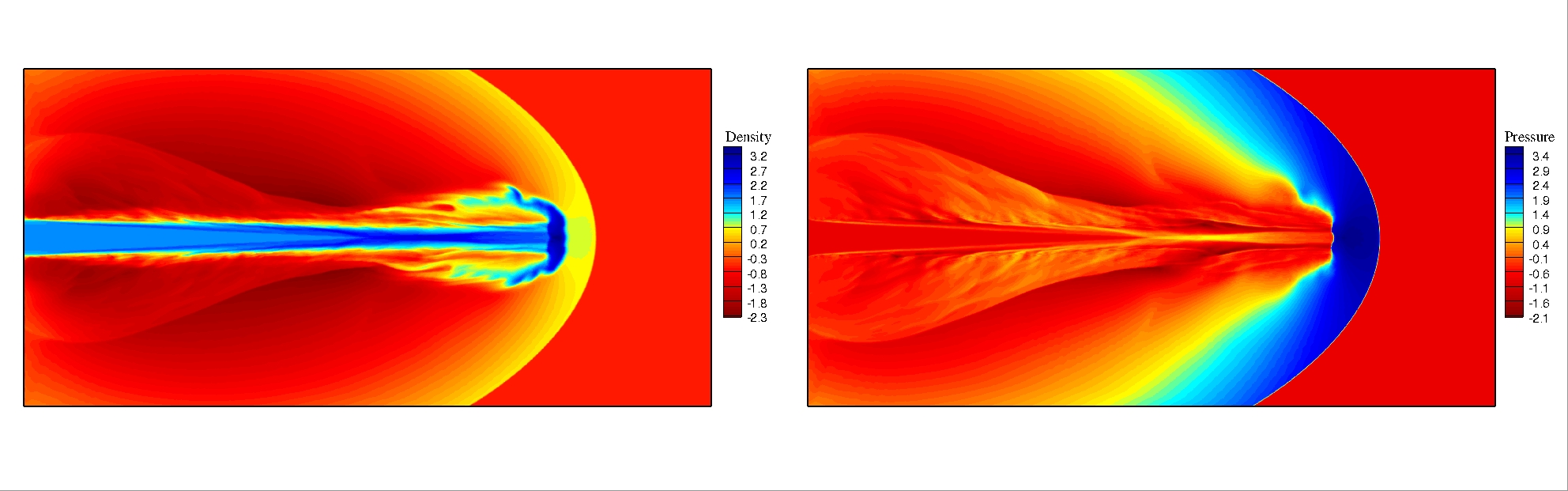}
\caption{Astrophysical jet problem: density (left) contours and pressure (right) contours of logarithmic scale using the $K_3-\tvd$ scheme with $1000\times 500$ grid points.}
\label{fig:jet}
\end{figure}

Astrophysical jets refer to high-speed gas flows with extremely high Mach numbers
that are captured by the Hubble Space Telescope. In this section, we consider 
a two-dimensional astrophysical jet without radioactive cooling \cite{ZHANG20108918}. 
From the computational point of view, this is a very challenging case since the extremely 
high kinetic energy may lead to negative internal energy.\\
The computational domain is $[0,2] \times [-0.5,0.5]$. The initial conditions are
\begin{align}
&(\rho,u,v,p)\\
&=\left\{
  \begin{tabular}{lr}
  $(5,11,0,0.4127)$  & if $x=0$ and $-0.05\leqslant y \leqslant 0.05$, \\
    $(0.5,0,0.4127)$ & otherwise,
\end{tabular}
\right.
\end{align}
which leads to Mach 30 with respect to the cold jet.
The computed density and pressure contours are illustrated in Fig. \ref{fig:jet} using the TVD scheme (temperature used as the determinant),
where the bow shock is visible propagating into the ambient medium. It can be noticed 
that the instabilities that appear around the jet are captured.
% \begin{figure}
% \centering
% \includegraphics[clip, trim=55cm 35cm 55cm 26cm, width=\linewidth]{jet-density-270}
% \caption{Astrophysical jet problem: Density contour.}
% \label{fig:compressible/jet-density-270}
% \end{figure}
% \begin{figure}
% \centering
% \includegraphics[clip, trim=55cm 35cm 55cm 25.5cm, width=\linewidth]{jet-Mach-270}
% \caption{Astrophysical jet problem: Mach-number contour.}
% \label{fig:compressible/jet-mach-270}
% \end{figure}
% Figure  \ref{fig:compressible/jet-mach-270} shows the Mach number contour in the domain,
% where the Mach number reaches 270 at the core of the jet.
It is expected that augmenting the model
with a positivity preserving limiter would lead to stable simulations for higher Mach numbers 
\cite{ZHANG20108918,zhang2011positivity,zhang2012positivity,fu2019very}.
It must be commented that the WENO scheme failed this simulation.\\

\section{Conclusion}
In this paper, we presented simulations of compressible flows in the PonD framework.
Using the shock capturing schemes such as WENO and TVD, we were
able to implement simulations in a wide range of Mach numbers; from mild
cases such as sod shock-tube to astrophysical jets. Comparison between the 
two numerical schemes were presented at each benchmark. 

The results show that 
the PonD model is able to handle highly supersonic flows when augmented
with proper numerical schemes. 
Moreover, it was observed that the TVD scheme features better performance in terms of accuracy and mass conservation.
However, for extreme cases, the model must be
equipped with more sophisticated techniques such as the positivity preserving schemes.\\

% Finally, we presented a conservative PonD formulation which preserves the total mass and energy during
% the simulations. The idea is based on monitoring the contribution of each point during the interpolations and imposing conservation by fixing the contribution of each individual point. The assessment of the proposed scheme was made through the simulation of Kelvin-Helmholtz instability, where the results illustrated the proposed scheme is able to conserve mass and total energy during the simulations. \\
\section{acknowledgments}

This work was supported by the European Research Council (ERC) Advanced Grant No. 834763-PonD and the SNF Grant No. 200021-172640 (E.R.). Computational resources at the Swiss National Super Computing Center (CSCS) were provided under Grants No. s897 and No. s1066.

% \appendix

 \bibliographystyle{elsarticle-num} 
 \bibliography{main}





% \end{thebibliography}
\end{document}